# A longitudinal study of field emission in CEBAF's SRF cavities 1995-2015

Jay Benesch
Thomas Jefferson National Accelerator Facility (Jefferson Lab)

**Abstract**

Field emission is one of the key issues in superconducting RF. When present, it limits operating gradient directly or via induced heat load at 2K. In order to minimize particulate contamination of and thus field emission in the CEBAF SRF cavities during assembly, a ceramic RF window was placed very close to the accelerating cavity proper. As an unintended consequence of this, it has been possible to monitor and model field emission in the CEBAF cavities since in-tunnel operation began. The ceramic is charged by field emission to a stable voltage and then discharges. This phenomenon had to be studied statistically to minimize the number of interruptions to accelerator operation for nuclear physics. We report here the results of our twenty year study of this and related phenomena.

## Monitoring Field Emission in CEBAF

The CEBAF cavity pair and helium vessel are shown schematically in figure 1. The features of interest for this work are the high resistance ($> 10^{12}$ ohms/square) cold ceramic RF window 7.62 cm from the beam axis, a fundamental power (input) coupler (FPC) with substantial magnetic dipole field, and sensors attached to the waveguide at room temperature. The FPC induces a transverse kick of ~20 milliradians–MeV/c when the electron is on-crest in the adjacent accelerating cell, $147^o$ away, and the cavity gradient as a whole is set at 7 MV/m.[1] While little trajectory modeling has been done[2], it is clear conceptually and has been demonstrated in vertical test dewar experiments that field emitted electrons from either cavity in a pair can reach and accumulate on the cold ceramic window.[3,4] The same set of vertical dewar experiments demonstrated that the interposing of an elbow or dogleg waveguide between the fundamental power coupler flange and the ceramic window dropped the electron current to the window by three orders of magnitude.

During CEBAF commissioning, arc discharges were seen at the cold ceramic windows via photomultipliers and vacuum sensors attached to the warm-to-cold transition waveguide. These were verified with spectroscopic observation in vertical dewar tests[5] to occur at the ceramic and may be either surface flashover or punch-through. The latter is demonstrated by leak testing – most of the cold ceramic windows in the accelerator now have holes in them. In vertical tests, with a picoammeter available to monitor field emission current to the window, the discharges occurred at roughly constant charge.[4] There is no way to monitor field emission current directly in the accelerator as all the vacuum seals are metal and there is no direct access to the cold ceramic window. All that can be recorded is the incidence of arc and vacuum faults and the gradient in the cavity at the time each fault occurred. Such records have been maintained since January 30, 1995. The data analyzed covers ~5700 cavity-years in tunnel.

This analysis assumes that the cold ceramic window is a perfect capacitor and that the charge at which a discharge occurs is constant. The interval between discharges is then inversely proportional to a constant field emission current. If the cavity gradient is constant throughout the interval and the RF is on throughout, one can easily apply a simple exponential or more rigorous Fowler-Nordheim[6,7] model to the data directly to obtain a field emission model for each cavity.

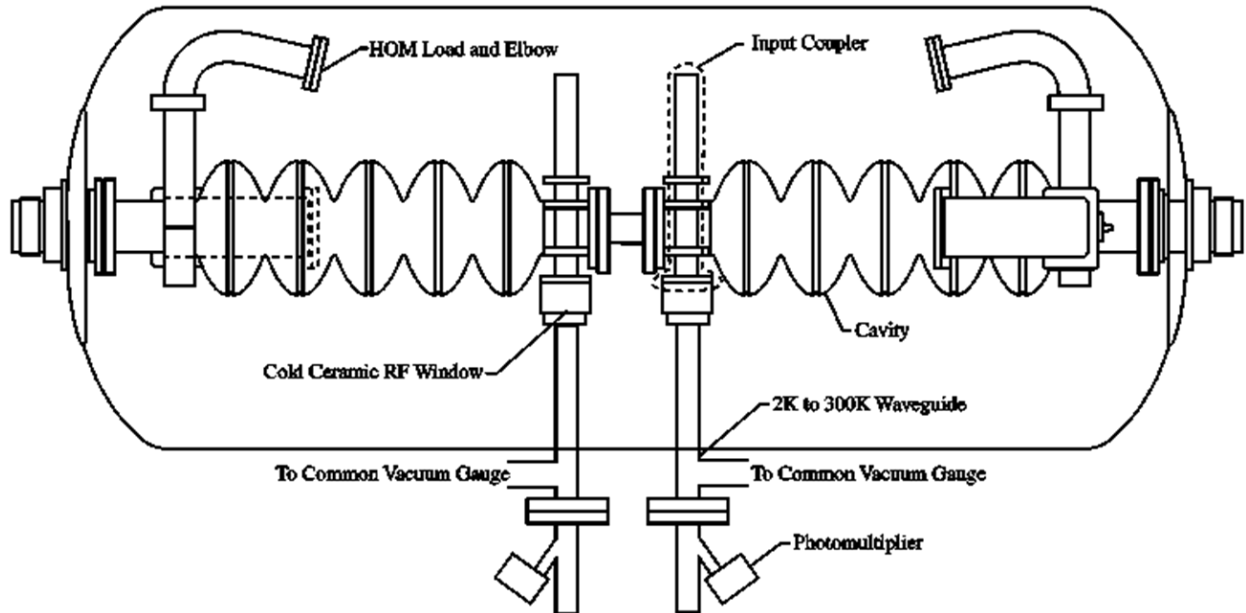

Figure 1. Cavity pair in helium vessel and room temperature sensors used in study.

The data is not perfectly clean, of course, so one pre-processing step and five data cuts are applied before statistical analysis. Until November 2004, there was no recordable signal giving RF-on time for each cavity. The pre-processing step approximated RF-on status by removing periods of 6+ hours in which no fault occurs anywhere in the machine from a running total of elapsed seconds. This assumes that all cavities are turned on and off at the same time, which is not the case – often one linac is on and the other off. This increases the noise in the data. The five cuts in the data and their justification are:

a. exclusion from analysis of faults with gradient under 3 MV/m due to limitations in RF control system stability which decrease fault interval
b. exclusion from analysis of faults with intervals under 30 seconds due to variation in reset time from 7-30 seconds; reset was manual during the first third of the data collection
c. exclusion from analysis of faults with intervals more than 12 days due to data plots suggesting that the assumption of perfect capacitors begins to break down at this interval. 12 days = 1036800 s. Data analyzed thus spans 4.5 orders of magnitude in interval.
d. exclusion from analysis of faults in which the gradient change from the preceding fault is more than 15%. There would be insufficient data to analyze if the assumption that the gradient is constant across the full interval were rigorously enforced. Both 10% and 15% cuts have been used with little difference in results. Since the gradient enters in the first power in the exponent in the simple exponential model and as the 5/2 power in the exponent in the Fowler-Nordheim model, no larger allowances were tested.
e. exclusion of simultaneous (within timing resolution) faults of cavities in multiple helium vessels as due to beam strike or control system effects rather than field emission.

Photomultiplier and vacuum sensors were mentioned above. The first is termed the arc detector and is a simple threshold detector – if a PMT signal greater than a fixed level is detected for more than 0.5 ms, the RF is shut off and a fault bit set. The second is connected to a pair of cavities and the actual pressure archived as a function of time. About 20% of the faults show only vacuum faults and cannot be assigned to a single cavity, only to a pair. Inclusion of these faults in the analysis of either member of the pair has always decreased correlation coefficients, so these faults are discarded. Some fraction of these are likely accompanied by sub-threshold PMT signals and should be included but there is no obvious way to determine which. About 75% of the faults show simultaneous arc and vacuum faults.

About 5% of the faults show only an arc detector bit. In early 2003 the archiving rate for the vacuum data was increased to 10 Hz. This allowed the addition of a pre-processing step which determines if there was a sub-threshold vacuum event at the same time as the arc detector fault and reclassifies about half of these 5% as true arcs. An increase in vacuum reading at least equal to background is required for the reclassification. When plotted all such vacuum traces show classic burst and recovery patterns.

The analysis which follows therefore begins with about 77% of the faults recorded and makes the cuts described above to this subset, ending with ~71% of the total faults. Known noise sources include, as discussed above: imprecision in RF-on intervals, changes in gradient during intervals, variation in window charge at discharge, and nonassignable vacuum-only faults.

Figure 2 below is the first cavity in CEBAF which varies in gradient; the two preceding cavities take the beam from 0.5 MeV kinetic energy to 5 MeV/c momentum and are invariant. Only seven points were removed by the data cuts discussed above. Residuals of the fit are shown in figure 3 left. It is close to normal visually but does not satisfy the Shapiro-Wilk W test for normality. Removing two outliers from the high side and eight from the low, followed by refitting, results in the residual distribution in figure 3 right, which is consistent with normality.

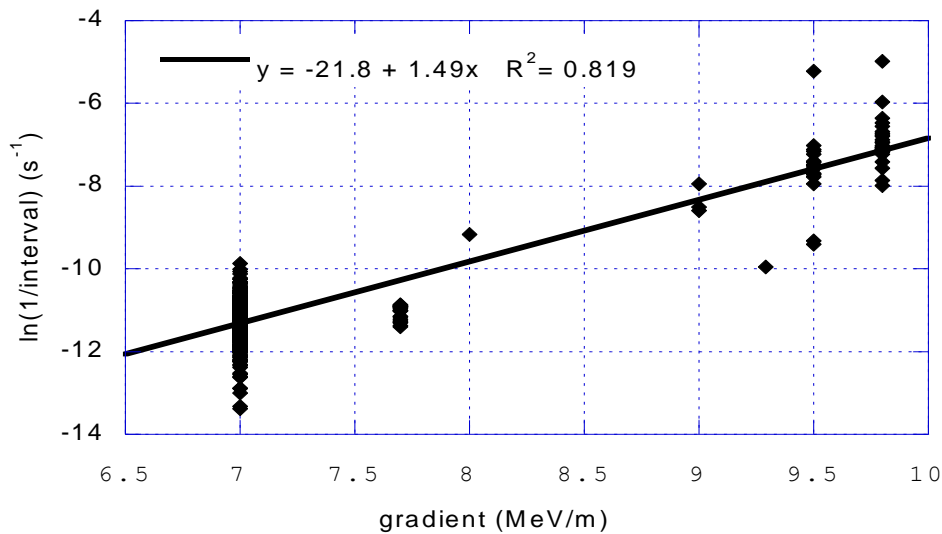

Figure 2. Fit to 0L031 data, automatic cuts only

This labor-intensive process of exploration of the data sets for outliers and development of exponential and Fowler-Nordheim models for each of 338 CEBAF cavities has been repeated many times since the beginning of 1995. The exponential models are used in a program which sets the gradient distribution along the linacs to minimize arc rate.[8] The Fowler-Nordheim models were used through August 2003 during outlier removal as residuals tended to be closer to normal (Shapiro-Wilk W test). After a hurricane-induced temperature cycle to room temperature in August 2003 such niceties were abandoned due to lack of time and only the exponential models were developed since those are the only ones used in machine setup.

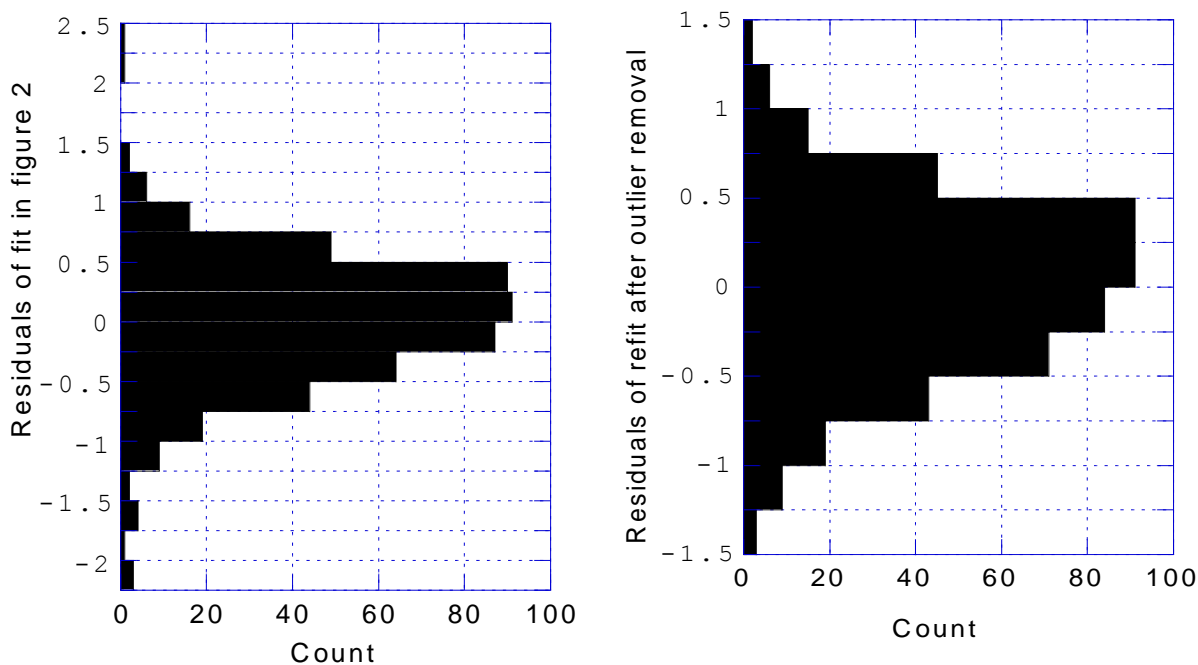

Figure 3. Residuals of fit in figure 2 before (left) and after (right) outlier removal.

There were 424867 faults with simultaneous light (arc detector) and waveguide vacuum excursions January 30, 1995 through May 12, 2012. There were 240 such faults in 2013 and 16537 in 2014, with all of the former and 6588 of the latter recorded during test or maintenance periods. The cavities were at 300K for eleven months between the two periods so the accelerator could be upgraded for higher energy operation. The raw and processed data can be made available to other investigators. The raw data includes every turn-on and turn-off of an RF system, the paired vacuum faults, beam strikes, RF drive system issues; seventeen indicators in total.

Two analysis tools are used during machine operation to maintain the statistical models needed to set machine gradients.[8] Since 2003 most of the work is done with a routine written in R[9] which applies four variations on least squares fitting to the data as in figure 2, plots the data and fits, and writes the fit equations to a text file in a format designed for spreadsheet import. The second, which is used when the automated fit plots or fit parameters raise questions in the author's mind, is a commercial data exploration pacakge, JMP[10].

We return to cavity 0L03-1 to use it for a more detailed analysis example. The data set used here is almost a decade long, October 2003 through May 2012 while that in figures 2 and 3 encompasses only the first two years. For electron optics reasons this cavity is generally set at 7 MV/m, 73% of the faults are at that gradient. Figure 4 shows the R output.

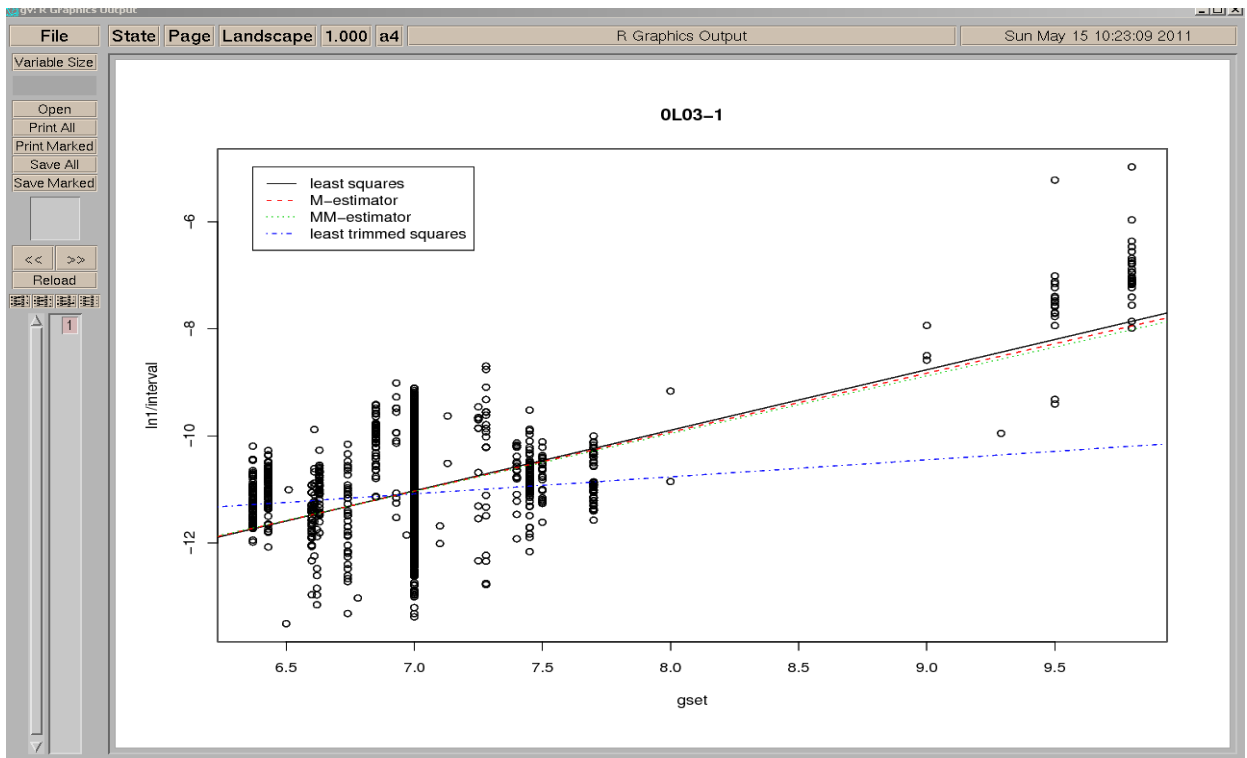

Figure 4. R program output. Three regression techniques agree; the fourth and most aggressive in trimming outliers does not. Such disagreement is an indicator that more investigation is warranted.

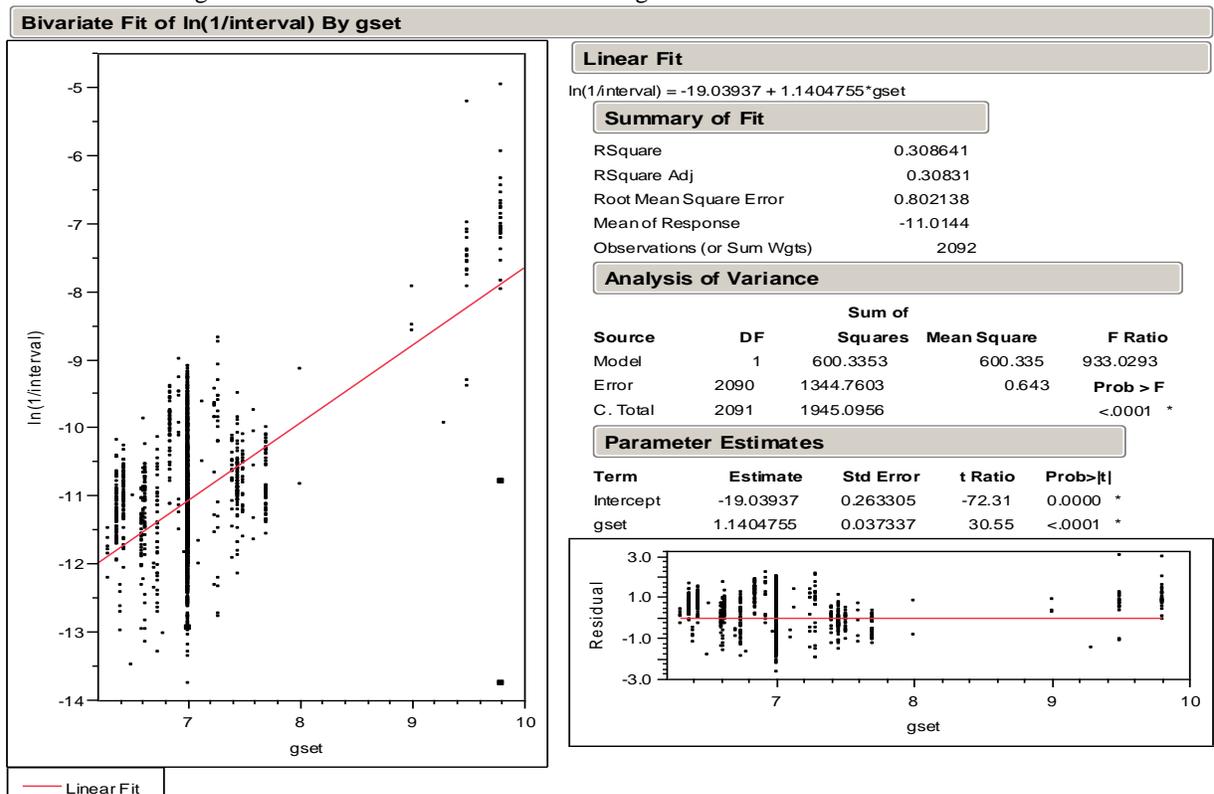

Figure 5. Same 0L031 data plotted in JMP. Note the "t ratio" = Estimate/(standard error). The fit is quite significant in spite of the large span (~$e^4$) of intervals at the dominant 7 MV/m.

One may apply the statistical technique *stepwise regression* to determine if any of the other cavities in the cryomodule has an influence on the fault interval in this one. Only two are over 5σ: cavities 1 and 6, with t ratios 28 and 18 respectively for slope. Adding cavity 6 to the regression increases the correlation coefficient $R^2$ from the 0.31 shown in figure 5 to 0.46. Fitting the cavity 1 fault intervals to cavity 6 gradient produces the model ln(1/interval)= -11.7 + 0.79*cav6grad. If one wants one fault per day in cavity1 due to cavity 6 one must set the latter to 3.5 MV/m, unacceptably low. At 9 MV/m the fault interval is 16 hours, tolerable. This value was used in lieu of that suggested by the model for faults in cavity 6 driven by cavity 6 gradient. I term this phenomenon fratricide.

For this review all 424867 true arc faults (1995-2012) were re-examined in JMP. Many discontinuities in response were noted as they occurred during the seventeen years. These were verified or discarded. Other discontinuities were discovered anew in the data. The data was ultimately divided into 1922 time blocks for 338 cavities. Unintended changes occurred roughly once every 14 years per cavity. Another analyst would likely arrive at a different division because some of the changes are subtle. This report covers only the gross changes which any analyst would find. Of these 1922 blocks, 1569 had enough data for statistical models.

The 2013-2014 data was analyzed with the R-based tool except where disagreement among the four least squares methods indicated an issue. Analysis with JMP reproduces the R result for ordinary least squares (OLS) if one makes only the cuts specified above. Additional outliers obvious to a human but not to the simple cuts specified are discarded when using JMP, so the resulting fits differ slightly from those output by R. In what follows, when comparisons are made between periods, the model ensembles are either JMP-JMP or R/JMP(~90/10)-R/JMP to reduce systematic error.

**Abrupt Changes in Field Emission**

There were 410 instances of abrupt change in cavity performance during the period 1995-2012 exclusive of intended action and hurricane Isabel. Some 265 of these occurred across a maintenance period with no intended changes to the cavity. The balance, 145, occurred while beam was being delivered. In figure 6 we show a change in cavity performance during beam delivery.

Another system implemented with good intentions during CEBAF construction, which remains in place to this day, closes gate valves on either end of each cryomodule every time an entry is made to the tunnel. Closing and opening a gate valve generates particulates which may travel meters through a cryomodule.[11] The author hypotheses that many of the changes in cavity performance are a result of these particles. There is no statistically significant correlation between cavity location within a cryomodule and performance changes in the data set. This does not reject the hypothesis because a clean room test done with gate valve and air-borne particle sensor in 1993 showed 2-3 meters of propagation in air. One would expect longer distances in vacuum. Figure 7 shows a change which occurred across a maintenance period. Another possible source of particles is the discharges at the cold ceramic windows. Many of these punch through the windows and could create a small cloud of particles in the cavity.

Cycling a cryomodule to ~35K while pumping on the beam tube removes adsorbed helium and hydrogen. This often improves performance. It may be that the 145 changes in field emission which occurred during beam operation were a result of adsorption of gas "sharpening" an existing asperity or changing the local work function sufficiently to raise the local electric field above the field emission threshold.[12-14]

The changes in performance for the two sets are quite different. The 145 changes during runs cause a mean loss of 15% in cavity performance, figure 8. The 265 changes across maintenance periods show a mean improvement of 6%, figure 9. The author has no hypothesis to explain this.

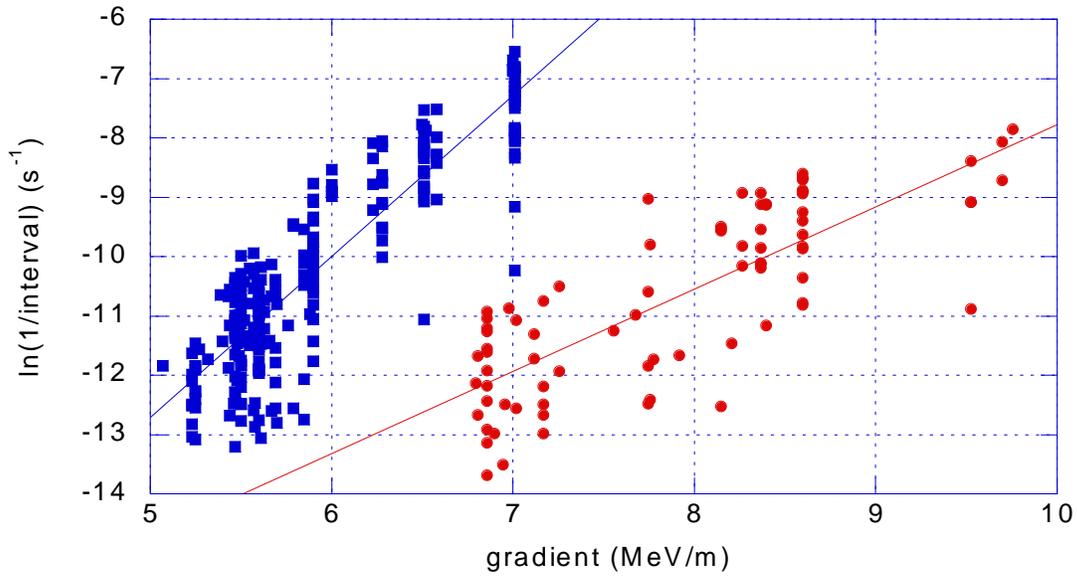

Figure 6. Blue points are after 0440 9/21/2004. Interval at 8.1 MV/m changed from ~80,000 seconds to ~500 seconds. Linear fits for the data sets before and after are shown. Cavity 2L145. No tunnel entry.

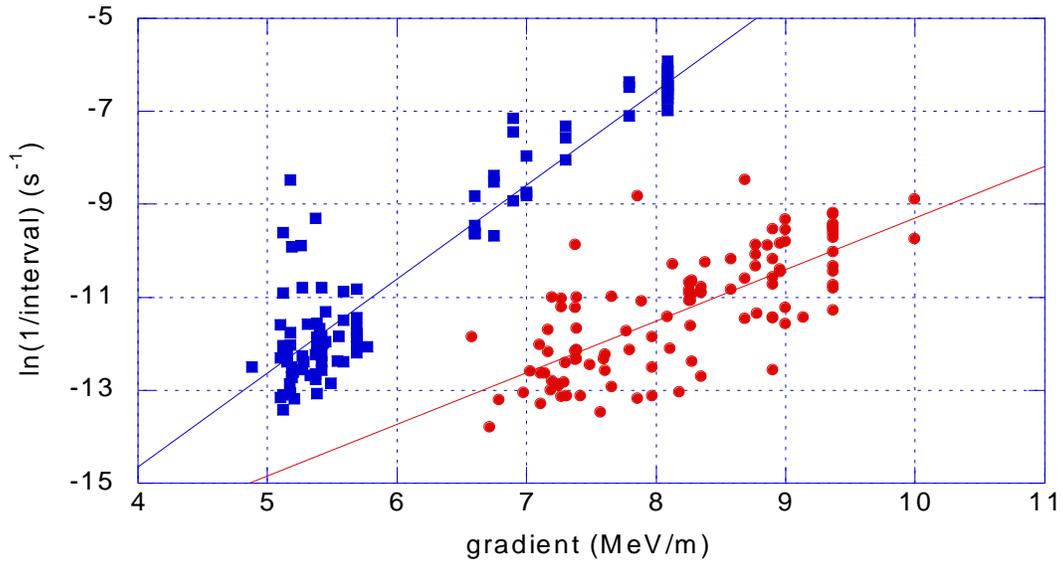

Figure 7. Data (blue) sloping across left of figure is after 5/18/2004. This change in behavior occurred across a maintenance period during which there was only an RF power cycle and a gate valve cycle. Cavity 1L087.

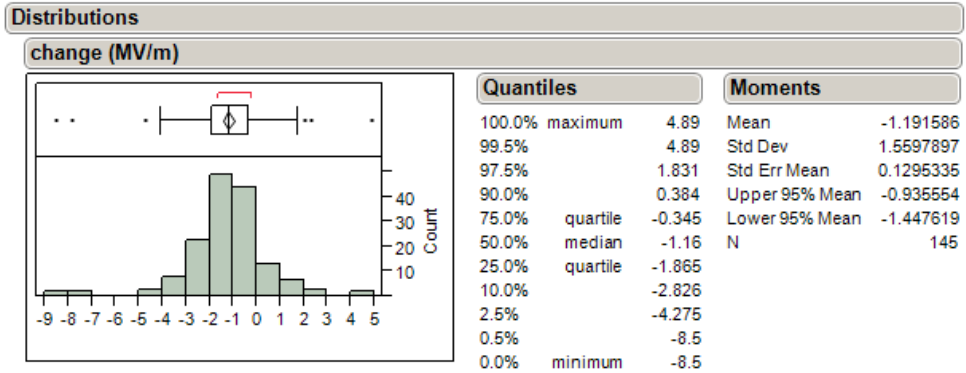

Figure 8. Change in gradient for two day fault interval due to abrupt changes in cavity performance during beam operation, like the example in figure 6.

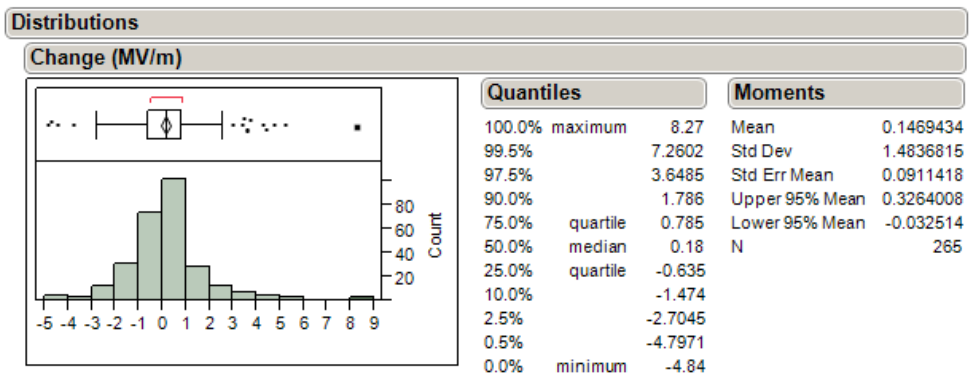

Figure 9. Change in gradient for two day fault interval which occurred across maintenance periods without work on the cavities in question, like the example in figure 7.

**Another phenomenon of interest: fratricide**

In addition to field emission in a cavity charging its own ceramic window and causing arcs, it is possible for an adjacent cavity to do so as well. This was mentioned above with respect to 0L031. The best example of this is cavity 2L06-4. The gradient in 2L06-5, in the following cavity pair, has a striking effect on it as shown in figure 10. If gradient in cavity 5 is below 9 MV/m, the usual own-cavity gradient dependence is obtained for cavity 4. The gradient in cavity 2L06-5 was limited to 8.5 MV/m as a result of this analysis.

Most at Jefferson Lab, including the author, were reluctant to believe in the effect. The author was convinced by the abrupt changes in the performance of cavities 6 and 7 in zone NL04 when an accidental introduction of $N_2$ into cavity 8 forced its gradient to drop from 10 MV/m to 5 MV/m. Retrospective analysis of previously misunderstood data showed that when cavity 8 was below 7.5 MV/m, fault vs gradient behavior in cavities 6 and 7 was consistent with field emission models. The physical mechanism of interaction between cavities which are not in the same pair is unknown. When fratricide is statistically found in cavity response and the culprit is located, a maximum culprit gradient is estimated and tested in the machine.

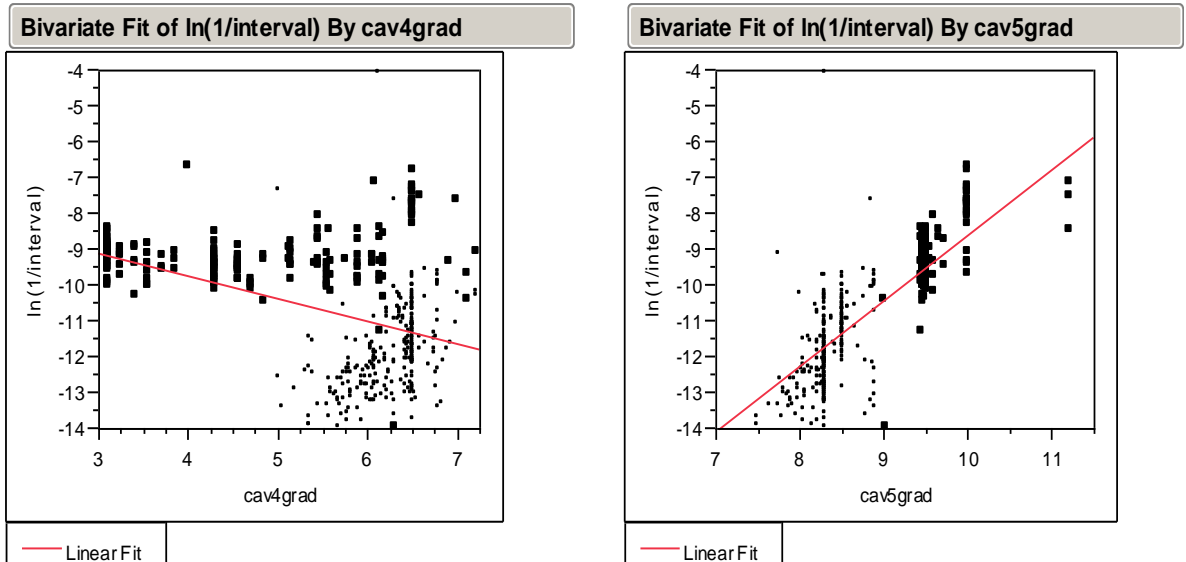

Figure 10 left. Fit of cavity 4 intervals by cavity 4 gradient. Larger points have cavity 5 gradient > 9 MV/m. No gradient dependence is seen for these points. Gradient dependence is seen when cavity 5 < 9 MV/m, as in lower right quadrant of left plot.

Figure 10 right. Fit of cavity 4 intervals by cavity 5 gradient. $R^2$=0.62. Same points highlighted in both.

## Characteristics of full ensemble of models 1995-2012

Characteristics of the ensemble of statistical models obtained during the period of 4-6 GeV CEBAF operation will now be presented. All of these were derived by the author using JMP, starting with the data cuts described above. Perhaps the most striking feature is an exceptional correlation of the slope and intercept of the exponential models (figure 11). This is a common feature of field emission systems and a statistical model is given in reference (15).

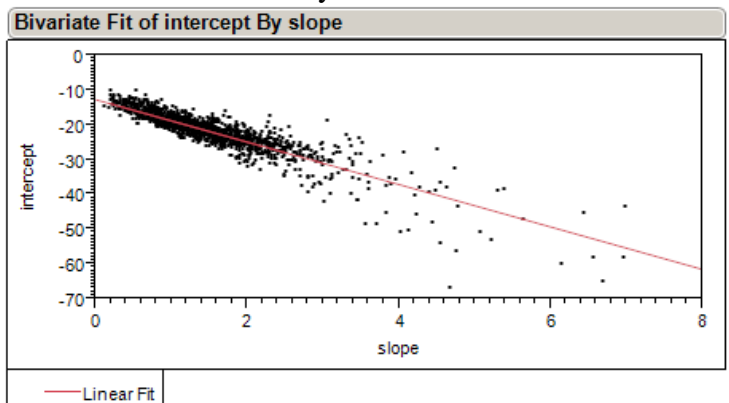

Figure 11. Plot of intercept vs slope. Fit is "Intercept = -12.63-6.10*slope". $R^2$ =0.88. t ratio of intercept estimate (-12.62) is 108.4; t ratio of slope coefficient is 107.9. 1569 models included, six outside plot.

The t ratio is (parameter estimate)/(standard error). Models with t_slope under 2, aka poorer than 95% exclusion of zero, were used in setting up the machine only in 1995 when data collection began and after prolonged periods at 300K when little data was available (see next section). The models are provided in a spreadsheet as supplemental online material. The models and their quality are summarized in figure 12.

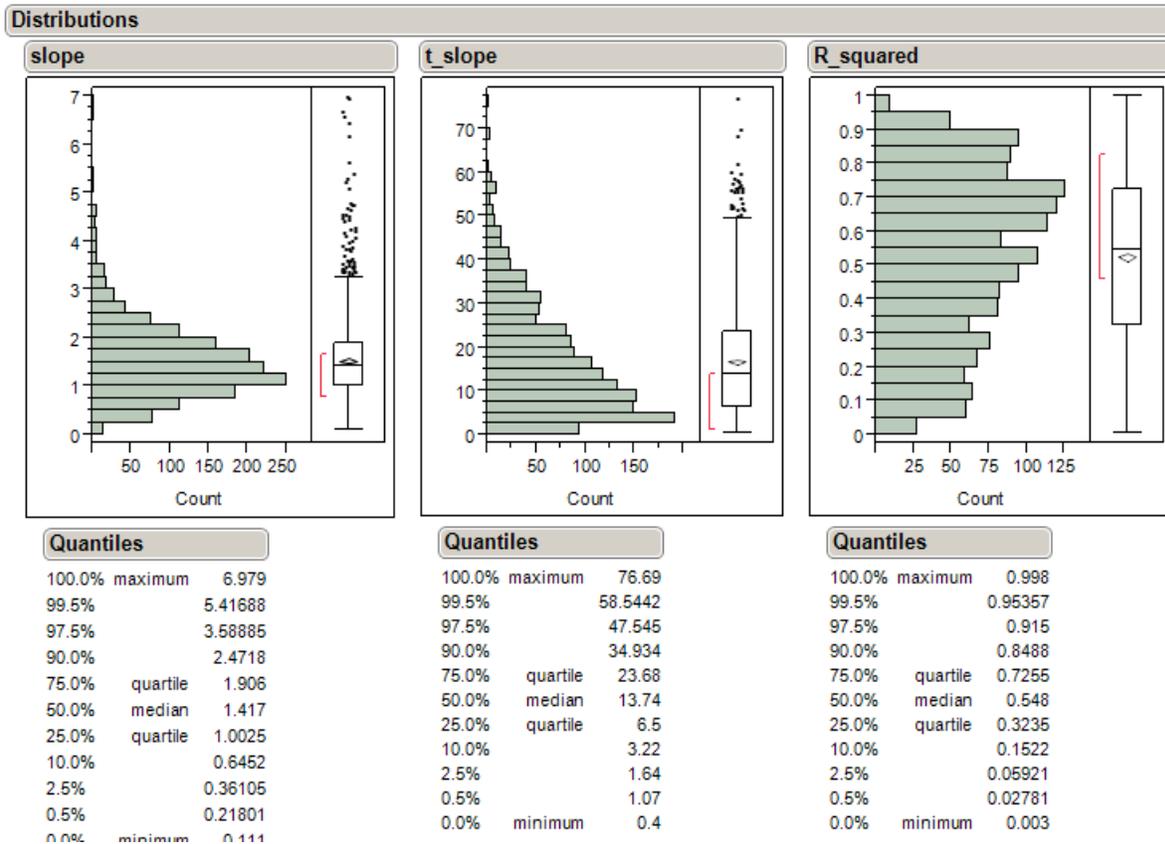

Figure 12. The left histogram shows 1561 model slopes for 1995-2012. Eight larger values are removed for clarity. The correlation in figure 11 shows that slope suffices to specify a model. Information on the quality of the 1561 models is given in the center and right histograms. t_intersept may be similarly estimated from t_slope,. t_intercept = -5.15-1.76*t_slope, $R^2 = 0.82$

**Effects of major accelerator perturbations**

Three events resulted in major perturbations of multiple cryomodules. In August 2003, hurricane Isabel caused a four day power outage and all cryomodules warmed to room temperature without control. In 2008 and 2009, maintenance of the main helium liquefier forced the cryo load to be shifted to a much smaller 4K refrigerator. The smaller capacity of this unit forced nine cryomodules to be warmed to room temperature. Third, in May 2012 the accelerator was shut down for a major energy upgrade and all modules brought to room temperature. Commissioning of the altered machine began in late 2013. In figure 13 gradients predicted to yield two day fault intervals are compared for March 2003 models, before the uncontrolled cycle to room temperature, and models after the incident. Figure 14 shows the smaller damage done with controlled thermal cycle to 300K in 2008 or 2009 due to helium cold box maintenance. Finally, Figure 15 shows the changes which occurred during the year the linacs were at 300K due to the recently completed CEBAF energy upgrade. The models used in constructing this figure differ from all those used above. They were constructed manually using JMP. The models used in figure 15 were made using the R process described above with manual intervention only when the four least squares routines differed sufficiently to warrant it. The models in figure 15 were used in accelerator setup. The 1569 JMP models otherwise discussed were created for this review.

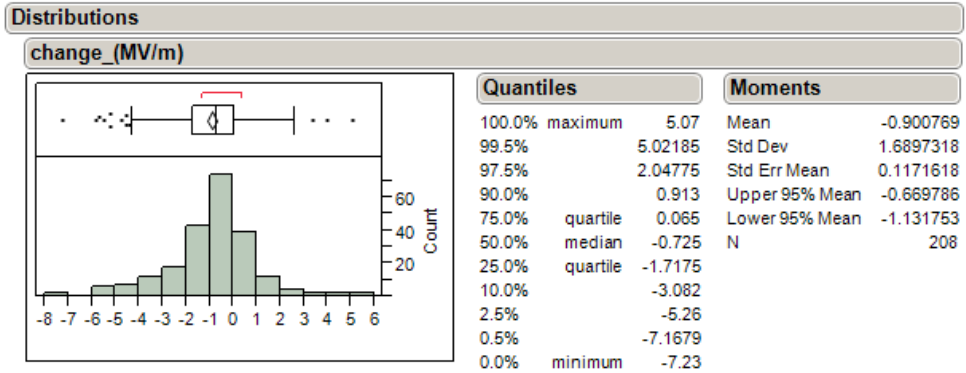

Figure 13. Distribution of changes in gradient expected to produce faults at two day interval across hurricane Isabel uncontrolled thermal cycle. Mean and median fractional change in gradient (not shown) was 10%.

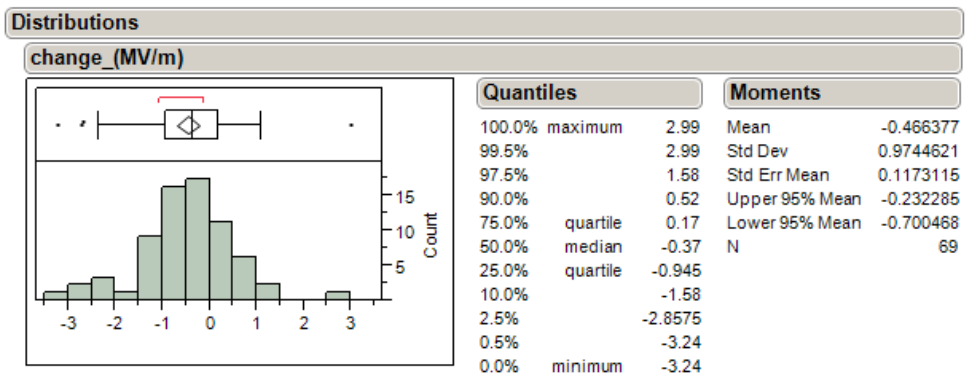

Figure 14. Distribution of changes in gradient expected to produce faults at two day interval across controlled thermal cycles in 2008 or 2009. Mean and median fractional change in gradient (not shown) were -6.3% and -5.4% respectively.

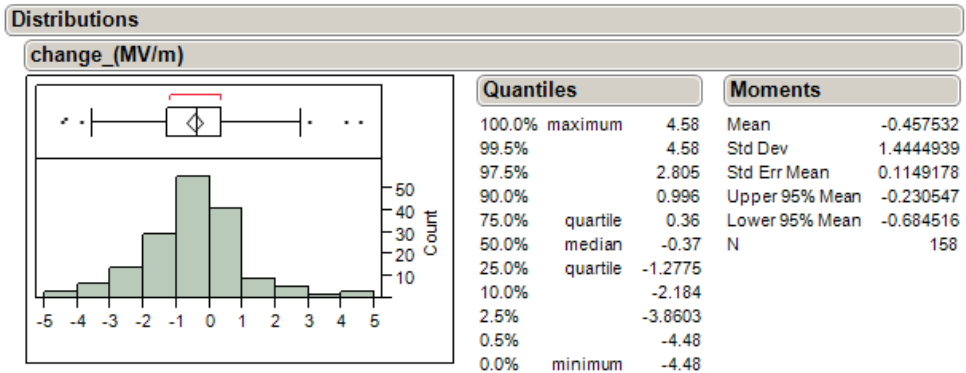

Figure 15. Distribution of changes in gradient expected to produce faults at two day interval across controlled thermal cycle which began May 2012 and lasted for one year. "After" data collected in 2014. The remarkable agreement between medians and means in figures 13 and 14 was not contrived. It may be a result of following the same written vacuum procedures at 300K to prepare for 4K cooldown. The other quantiles and the standard deviations differ substantially. Median fractional change -6%, mean fractional change -4.6%.

Based on figures 14 and 15, energy reach calculations for CEBAF include a 0.5 MV/m decrease for cavities when a 300K cycle is performed. Similar data must be collected at another accelerator to determine if the MV/m change or fractional change is most appropriate for others to use in estimating the effect of thermal cycles: 6% of 35 MV/m is four times 0.5 MV/m.

**Mitigation**

Two mitigation methods have been pursued. When funds were available 2007-2010, ten modules were refurbished. One more was refurbished in 2013. They were completely disassembled. The cold ceramic window was moved radially outward about 18 cm. A bowed niobium waveguide was placed between the cavity and the new ceramic window location to eliminate line of sight from window to cavity. Only two of the 80 cavities in these ten modules have shown any signs of window charging and the gradient reductions required to eliminate resulting RF faults were small.

The first mitigation pursued is helium processing[6]. At Jefferson Lab, the procedure once began with a cycle to ~35K to remove adsorbed helium and hydrogen, establishing initial conditions. After refilling the module and pumping back down to 2K, a controlled amount of helium is introduced to the cryomodule bore at one end, to produce a gas pressure of a few tens of microtorr within the eight cavities. Instrumentation appropriate for measurement of gaseous helium pressure in this range was NOT installed; pressure was established by connecting a known small volume near atmospheric pressure to the cryomodule and adjacent warm magnet assembly. Eight or ten small Geiger-Muller tubes are attached to the exterior of the module; the two on the beam pipes at each end generally would have sufficed. RF was established in one cavity and gradient increased until a substantial radiation reading was obtained. Over tens of minutes the reading would decrease and the gradient would be increased. It is hypothesized that back-bombardment of the field emitter by helium ions "blunts" the emitter, lowering local electric field and emitted current. Alternatively[12-14], adsorbed gas may be moved, increasing the local work function. Multiple field emitters may be found sequentially in a cavity as the gradient is increased. When available RF power is exhausted or there is no further change in radiation level with power, the process is stopped on that cavity and begun on the next.

There is debate among Jefferson Lab staff whether there is something special about the first round of helium processing versus subsequent rounds, aka diminishing returns. Figures 16 and 17 show improvement due to first and subsequent helium processing respectively. Mean improvement was 41% on first application and 31% on subsequent application. Some of the 10% reduction in relative improvement is thought due to proximity in time between applications - new emitters didn't have time to develop.

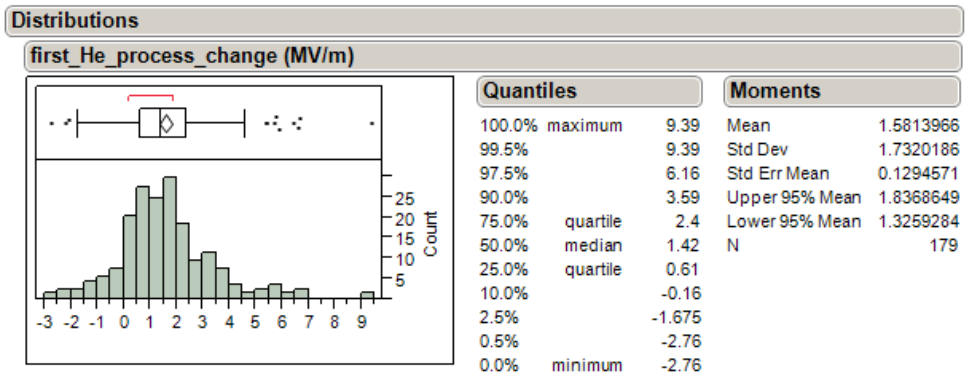

Figure 16. Improvement in gradient for two day fault interval after first application of helium processing. If no model was available either before or after processing, for instance if no faults were observed at maximum RF power available, no difference can be calculated.

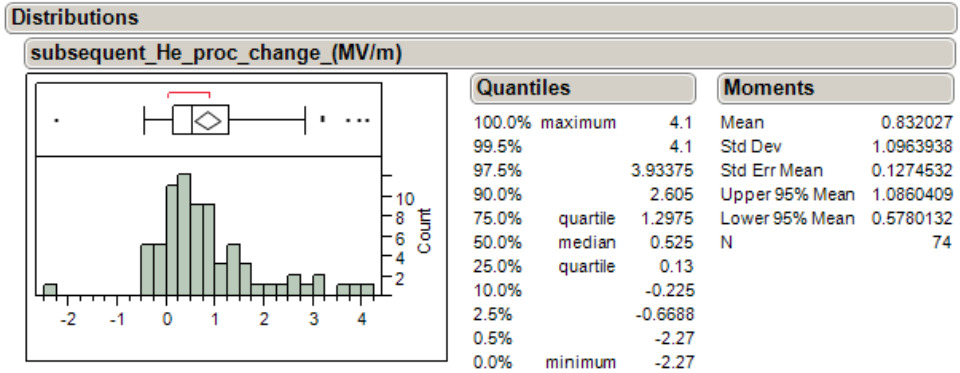

Figure 17. Improvement in gradient for two day fault interval after subsequent applications of helium processing. Again, there are cases which lack either "before" or "after" models so no differences can be calculated.

Based on Figure 17, helium processing of all CEBAF cavities has been scheduled for 2015. Even refurbished modules are often limited in performance due to heating produced by field emission, either to quench or to available 2K flow, and so are expected to profit by the processing as well. Performance improvement of cavities without the close cold ceramic windows and the unfortunate charging phenomenon here exploited will be difficult to quantify because CEBAF does not have x-ray monitors installed for routine operation. The G-M tubes used for helium processing are moved from zone to zone during the work. There are eleven refurbished modules with about 400 cavity-years for which no time series data is available.

During 2013 and early 2014 ten cryomodules of a new design were commissioned in the tunnel. The movable GM tube array described above was used to monitor field emission during this effort. Gradients were calibrated calorimetrically and with RF equipment; beam time was not available in 2014 to do an adequate job of measuring the momentum gain from each. In figure 18 the field emission onset gradients during commissioning are displayed.[16] The 25 MV/m upper bound is an administrative limit. Design gradient for operation with beam is 18 MV/m.

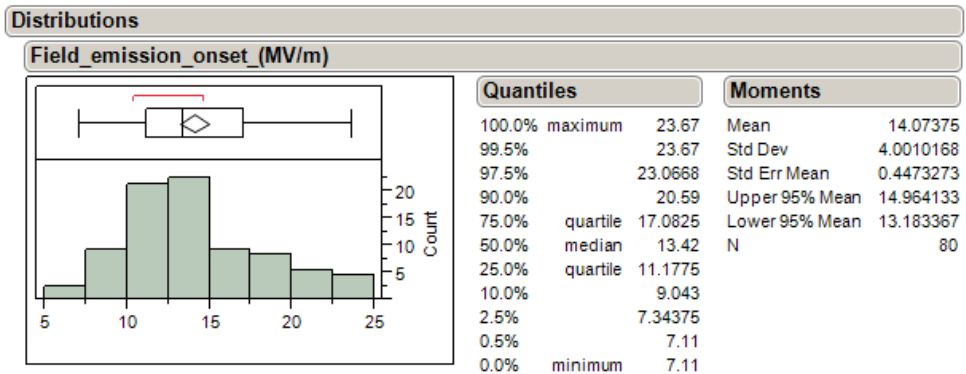

Figure 18. Field emission onset gradients during C100 cryomodule commissioning in CEBAF

**Implications for the future accelerators**

Hypotheses related to the abrupt changes in field emission behavior include adsorbed gas sharpening an existing asperity, adsorbed gas changing the local work function, and particle motion. The very different structure of the vacuum system in CEBAF, individual cryomodules, versus those in the X-FEL and planned for LCLS-II and ILC, will reduce particulate and adsorbed gas sources: very few gate valves, no ion pumps and no warm girders.  The fact that CEBAF runs CW with RF on ~75% of the year and the ILC will run pulsed with ~1% duty cycle should be irrelevant as vacuum effects almost certainly govern, not RF.

CEBAF runs with ~600W of 2K heating due to field emission at 5.8 GeV, or ~2.5W per cavity with field emission model.  In vertical dewar tests, the author was able to run with up to 70W of field emission heating without quenching the cavity.  Maximum field emission heat load allowable by the LCLS-II and ILC cryomodule designs are not known to the author.  Some allowance must be made.

Figure 18 shows that assembly improvements since original CEBAF construction have not adequately reduced field emission when there is a 2K to 300K transition at each end of each cryomodule.  Cavities must be set individually to maximize momentum gain within each cryomodule subject to the liquid helium flow restrictions in the system (aka 250W).  It will be useful to watch LCLS-II performance, comparing modules assembled at JLab and FNAL.

Field emission in cavities with high gradients and total length of order one meter produces electrons with energies well above the 5 MeV necessary to excite the giant resonance in high Z cryostat materials and activate them.  This may become a gradient limiting issue in the new CEBAF cryomodules or prevent tunnel access in their vicinity and that of the first bending magnet following.  Use of low Z metals rather than stainless or carbon steel should be considered in future accelerators.  Field emission at any energy levels will harden and embrittle elastomers so vacuum valves should have metal seals.  These two constraints suggest that the use of gate valves be minimized, as does the number of particles they generate[11].

One possible mitigation system for the reported field emission change phenomenon in future accelerators would require:
1. x-ray detectors near the beam pipe every four cavities, half as cold spares
2. RF system capable of varying power to individual cavities
3. software to detect changes in x-ray patterns and use (2) to determine which cavity is at fault parasitically during normal operation.  Energy lock assumed.
4. Sufficient momentum headroom to allow for gradient reduction in individual cavities as new field emitters turn on.

**Gradient calibration**

Gradient calibration has been a problem in CEBAF because of the range of cavity performance due to field emission and the number of cavities in each linac, even with magnetic spectrometers, aka arcs, after each linac.  Calibrating the highest gradient cavity in the linac magnetically and then using it as a reference for making null measurements against all other cavities is subject to the lower bound of low level control RF (LLRF) system stability and therefore has a sample standard deviation of 7%.  The standard deviation of changes from RF to beam-based calibration was 16%.  Since there are eight cavities in a cryomodule, the 7% result was tolerable for 6 GeV CEBAF.  With the emittance increase via synchrotron radiation due to the increase to 11 GeV in the five original passes, greater accuracy is needed.  Use of the phase shift capacity in the LLRF system will allow for larger momentum swings for poor cavities, allowing a reduction in error of perhaps a factor of three.  This will be tried in November 2015.  It is required for accurate setting of the FODO lattice for the upgraded machine if beam quality sufficient for parity experiments is to be delivered.  Gradient calibration is not discussed in the ILC reference design report; I haven't checked the TDR.

**Summary**

Insights gained from two decades of monitoring and modeling field emission in CEBAF have been discussed.  Items possibly relevant to the future accelerators have been pointed out.  Most important, 2.5 sudden changes in field emission, yielding onset at substantially lower gradient, occur per cavity-century in CEBAF.  The phenomenon designated fratricide complicates diagnosis but can be dealt with using standard statistical techniques.  Monitoring of field emission via dedicated x-ray monitors in tunnel is desirable for future accelerators using superconducting RF.

**Acknowledgments**

Programming support for this work has been provided since 2002 by Michele Joyce.  Work supported by U.S. Department of Energy Office of Science Contract DE-AC05-84ER40150.  An unpublished earlier version of this work is http://arxiv.org/abs/physics/0606141.  Prof. R.G. Forbes of University of Surrey provided very helpful comments in email correspondence I initiated after seeing version 1 of reference 7 on arxiv.  He provided references 12-15.